\documentclass[aps,11pt,twocolumn]{revtex4}
\usepackage[dvipdf]{graphicx}

\begin{document}

\title{Ground state, electronic structure and magnetism of LaMnO$_3$}
\author{R.J. Radwanski}
\homepage{http://www.css-physics.edu.pl}
\email{sfradwan@cyf-kr.edu.pl}
\affiliation{Center for Solid State Physics, S$^{nt}$Filip 5, 31-150 Krakow, Poland,\\
Institute of Physics, Pedagogical University, 30-084 Krakow,
Poland}
\author{Z. Ropka}
\affiliation{Center for Solid State Physics, S$^{nt}$Filip 5,
31-150 Krakow, Poland}

\begin{abstract}
We have calculated the discrete low-energy electronic structure in
LaMnO$_3$ originating from the atomic-like states of the strongly
correlated 3\textit{d}$^4$ electronic system occurring in the
Mn$^{3+}$ ion. We take into account very strong intra-atomic
correlations, crystal field interactions and the intra-atomic
spin-orbit coupling. We calculated magnetic and paramagnetic state
of LaMnO$_{3}$ within the consistent description given by Quantum
Atomistic Solid State Theory (QUASST). Our studies indicate that
the intra-atomic spin-orbit coupling and the orbital magnetism are
indispensable for the physically adequate description of
electronic and magnetic properties of LaMnO$_3$.


\end{abstract}
\maketitle

The present study is a continuation of our systematic
investigations on the electronic structure and magnetic properties
of compounds containing transition-metal atoms. Problem of the
Mn$^{3+}$ ion containing compounds is of particular interest.
LaMnO$_{3}$ is an insulating antiferromagnet with the Neel
temperature of 140 K \cite{1,2,3,4}. Recently in LaMnO$_{3}$ the
problem of the orbital ordering is widely discussed. The orbital
ordering appears much above Neel temperature. For us the orbital
ordering is related to local distortions well accounted in terms
of crystal-field interactions. It is a reason for our long lasting
studies of crystal-field interactions in 3d-/4f-/5f-atom
containing compounds. The Mn ions in the perovskite structure sit
in the distorted oxygen octahedron. The basis for all theories is
the description of the Mn$^{3+}$ ion and its electronic structure.
There is no consensus about the description of this electronic
structure.

The aim of this paper is to present a consistent description of
the low-energy electronic structure and of the magnetism of
LaMnO$_{3}$. In our understanding, formulated as the Quantum
Atomistic Solid State Theory (QUASST) \cite{5,6}, the low-energy
electronic structure is related to the atomic-like electronic
structure of the strongly-correlated 3$d^{4}$ electronic system
occurring in the Mn$^{3+}$ ion. In our description the orbital
magnetism and the intra-atomic spin-orbit coupling play the
fundamentally important role.

The four $d$ electrons of the Mn$^{3+}$ ion in the incomplete 3$d$
shell in LaMnO$_{3}$ form the strongly correlated intra-atomic
3$d^{4}$ electron system. These strong correlations among the 3$d$
electrons we account for by two Hund's rules, that yield the
$^{5}D$ ground term, Fig. 1a. In the oxygen octahedron
surroundings, realized in the perovskite structure of LaMnO$_{3}$,
the $^{5}D$ term splits into the orbital doublet $^{5}E_{g}$ as
the ground subterm, Fig. 1b, and the excited orbital triplet
$^{5}$T$_{2g}$. For physically adequate description of states we
have to take into account the intra-atomic spin-orbit coupling
because it is always present in the ion.
\begin{figure}[!t]
\begin{center}
\includegraphics[width = 6.5 cm]{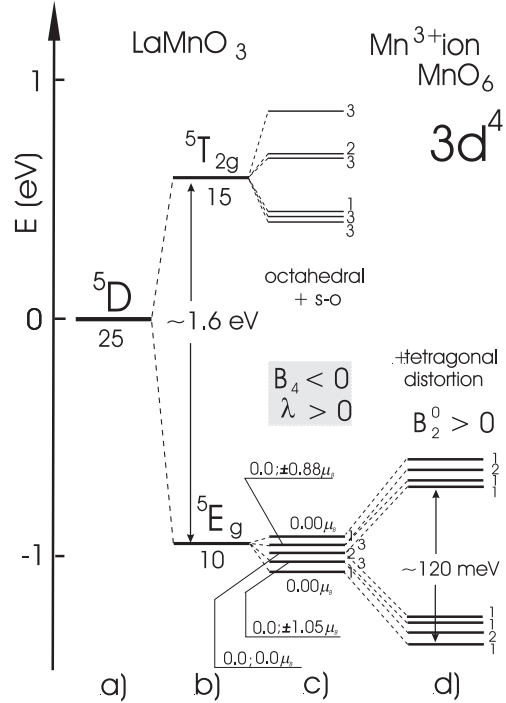}
\caption{Calculated electronic structure of the lowest $^{5}D$
term (a) of the 3$d^{4}$ electronic system occurring in the
Mn$^{3+}$ ion realized in LaMnO$_{3}$, produced by the octahedral
crystal field (b), the intra-atomic spin-orbit coupling (c) and
the tetragonal distortion (d). The degeneracy and the magnetic
moment of the states are shown. The fine splitting is not to the
left-hand scale.}
\end{center}
\end{figure}

The $E_{g}$ ground subterm comes out from \textit{ab initio}
calculations
for the octupolar potential, the A$_{4}$ CEF coefficient acting on the Mn$%
^{3+}$ ion from the oxygen negative charges. Such the atomic-like
3$d^{4}$ system interacts with the charge and spin surroundings in
the solid. The interaction with the charge surroundings we
approximate by means of the crystal-field interactions. As is
shown in Fig. 1c, the energy level scheme of the 3$d^{4}$ system
in the octahedral crystal field and in the presence of the
spin-orbit coupling contains 25 states in the spin-orbital space.
The dominating octahedral crystal field leaves 10 lowest states
well separated from others. These 10 states are split in two
quintets by the tetragonal distortion, see Fig. 1d. The splitting
of two quintets is of order of 120 meV and we think that this
splitting has been revealed in Raman spectra \cite{7}. This fine
electronic structure predominantly governs the electronic and
magnetic properties of real 3$d$-ion systems. To such electronic
structure we superimpose the spin-dependent interactions to
account for the macroscopically observed magnetic state. The
appearance of the magnetic state is associated with the spin
polarization and the temperature dependence of the energy of the
levels as is seen in Fig. 2. The self-consistent calculations have
been performed similarly to that presented in Ref. \cite{8}
for FeBr$_{2}$. With the octahedral CEF parameter $B_{4}$ = -13 meV, $%
B_{2}^{0}$=+10 meV, the spin-orbit coupling parameter $\lambda
_{s-o}$=+33 meV and the molecular-field coefficient n=26.3 T/$\mu
_{B}$ we get a value of 3.72 $\mu _{B}$ for the Mn$^{3+}$-ion
ordered magnetic moment. This value is in good agreement with the
experimental datum. LaMnO$_{3}$ has been calculated to order
magnetically along the $a$ axis within the tetragonal plane. The
magnetic interactions set up at 0 K the molecular field of 108 T.
The calculated magnetic moment of 3.72 $\mu _{B}$ is built up from
the spin moment of +3.96 $\mu _{B}$ and from the orbital moment of
-0.24 $\mu _{B}$. Although the orbital moment is relatively small
its presence modifies completely the electronic structure. The
Mn$^{3+}$ ground-state wave function is given as $\psi =
0.87|x^{2}-y^{2},-2> + 0.48|z^{2},-2>$.
\begin{figure}[!t]
\begin{center}
\includegraphics[width = 7.9 cm]{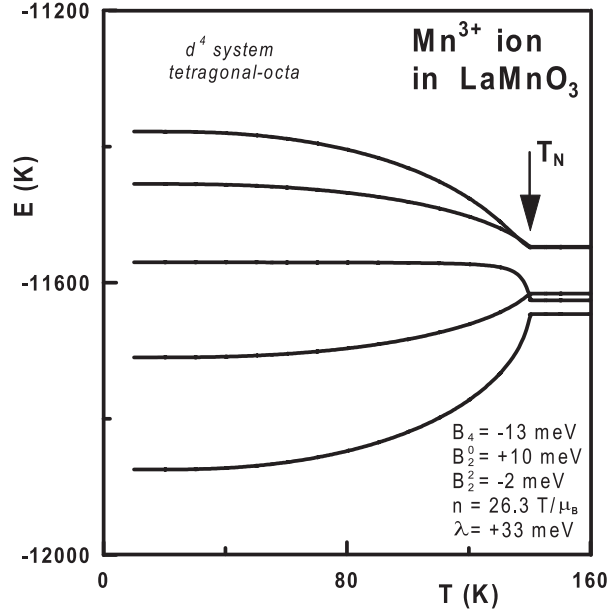}
\caption{Calculated temperature dependence of the energy of the 5
lowest levels of the Mn$^{3+}$ ion in LaMnO$_{3}$ in the
paramagnetic (T$>$T$_{N}$=140 K) and the magnetic (T$<$T$_{N}$)
state with the magnetic moments along the $a$ axis. Parameters are
in the figure.}
\end{center}
\end{figure}
Having the electronic structure, both in the magnetic and
paramagnetic state, we can calculate the free Helmholtz energy and
the resulting thermodynamical properties by means of the
statistical physics.

The calculated electronic structure differs fundamentally from
those presented in the current literature \cite{1,2,3,4} with
$t_{2g}$ and $e_{g}$ one-electron orbitals. In contrary to the
one-electron description used in presently-in-fashion theories we
work with many-electron functions of $S$=2 and $L$=2. In case of
the Mn$^{3+}$ ion, its four d electrons are put in literature
subsequently on the $t_{2g}$ (three electrons with spin up) and
$e_{g}$ (the fourth electron) orbitals. Such the picture we
consider as completely oversimplified - it neglects mutual
interactions within the incomplete 3d shell. In contrast, our
approach is the crystal-field approach with strong electron
correlations and with the importance of the spin-orbit coupling.
Comparison of different theoretical approaches, the band theory,
the one electron crystal-field approach and our many-electron
crystal-field theory has been presented in Ref. \cite{9}. We are
surprised that the presented by us many-electron crystal-field
approach is not popular among the 3d theoretists though it was Van
Vleck who already in a year of 1935, i.e. seventy years ago, has
shown its substantial applicability to iron-group compounds
\cite{10}. We continue the research by Van Vleck by taking into
account the spin-orbit coupling and by calculating
\emph{explicitely}, not by a perturbation method, the eigenstates
and eigenfunctions of the 3d paramagnetic ions in the Hund's rules
ground terms in a solid \cite{11}. Moreover, we manage to combine
the CEF theory with spin-dependent interactions, for Hamiltonian
see to Ref. \cite{8} - the same was used here for getting Fig. 2,
that enables to calculate the magnetically-ordered state within
one consistent procedure. Having eigenstates and eigenfunctions
all thermodynamics can be calculated. It means, that properties
like temperature dependence of the magnetic moment, of the
paramagnetic susceptibility and of the heat capacity can be
calculated from \emph{first principles}. Thus we were extremely
surprised that our paper showing a scheme similar to that shown in
Fig. 1 has been rejected by the Organizing Committee of
Strongly-Correlated Electron Conference in 2002, because for the
Committee and its referees it was unclear why the derived by us
the level structure with the $E_{g}$ ground state "is reversed
with respect to the $t_{2g}$-$e_{g}$ structure obtained by many
authors" \cite{12}. We could not believe that within the
Organizing Committee (and also within the International Committee
as we have asked to inform them about our protest) there was no
one who could understand our long explanations that the $E_{g}$
ground state comes out from \emph{ab initio} CEF calculations for
the Mn$^{3+}$ ion in the oxygen octahedron. Such a rejection
reveals the enormous shortage of knowledge about the CEF theory -
orbitals are not distinguished from the state symmetry. This
difference is clearly defined in text-books of Abragam and Bleaney
\cite{13}, Ballhausen \cite{14}, and many, many others. This
shortage of knowledge about the crystal field we met earlier in
Phys. Rev. B and Phys. Rev. Lett. Our papers from 1996 explaining
properties of LaMnO$_{3}$ with the $E_{g}$ ground state of the
Mn$^{3+}$ ion in LaMnO$_{3}$ (BZR586, BER639, LA6567 - it can be
found as Ref. \cite{15}) and pointing out the importance of the
spin-orbit coupling has been rejected as having an incorrect
ground state. Moreover, the referee wrote that our program must be
wrong as it splits the $E_{g}$ state. Although we could persuade
the referee that our program calculates well the splitting of the
$E_{g}$ state, the Editors have maintained the rejection. Thus, we
collect very large evidence for the discrimination of the
crystal-field approach with the spin-orbit coupling among the
solid-state community. It prohibits the open scientific discussion
about the magnetism and the electronic structure of 3d-atom
containing compounds. We are convinced, however, that the
scientific truth will finally come up. We claim that the
spin-orbit coupling and strong intra-atomic electron-correlations
have to be taken into account for the physically adequate
description. It is the highest time to "unquench" the orbital
moment in the 3d solid-state physics.

In conclusion, we argue that in LaMnO$_{3}$ there is preserved
low-energy electronic structure originating from atomic-like
states of the strongly correlated 3$d^{4}$ electronic system
occurring in the Mn$^{3+}$ ion. Our approach provides a
microscopic mechanism for the electron-lattice and the
spin-lattice coupling as well as for the electron-electron
correlations. We are convinced that the orbital degree of freedom
($L$) and the intra-atomic spin-orbit coupling are indispensable
for the physically adequate description of electronic and magnetic
properties of LaMnO$_{3}$. A discrete atomic-like electronic
structure in the meV scale has been revealed recently in
LaCoO$_{3}$ \cite{16,17} confirming the basic assumption of the
developed by us Quantum Atomistic Solid State Theory. QUASST
enables calculations of the value and the direction of the local
magnetic moment and reveals that they are predominantly governed
by the local lattice symmetry. We get a value of 3.72 $\mu _{B}$
for the Mn$^{3+}$ ion magnetic moment - this value depends on
local distortions and contains the substantial orbital moment. Our
approach accounts for the insulating state as well as for a number
of transitions observed in the Raman-scattering experiment on
LaMnO$_{3}$ and provides the proper symmetry of the ground state
eigenfunction.


\begin{thebibliography}{9}
\bibitem{1} Y. Tokura, N. Nagaosa, Science \textbf{288} (2000) 462.

\bibitem{2} V. S. Su, T. A. Kaplan, S. D. Mahani, and J. F. Harrison, Phys.
Rev. B \textbf{61} (2000) 1324.

\bibitem{3} Z. Popovic and S.Satpathy, Phys. Rev. Lett. \textbf{84} (2000)
1603.

\bibitem{4} E. Dagotto, T. Hotta, and A. Moreo, Physics Reports \textbf{344} (2001) 1-153.

\bibitem{5} R.J. Radwanski, R. Michalski, and Z. Ropka, Acta Phys. Pol. B
\textbf{31} (2000) 3079.

\bibitem{6} R.J. Radwanski and Z. Ropka, \emph{Quantum Atomistic Solid-State Theory}, cond-mat/0010081.

\bibitem{7} E. Saitoh, et al., Nature \textbf{410}
(2001) 180.

\bibitem{8} Z. Ropka, R. Michalski, R.J. Radwanski, Phys. Rev. B \textbf{63} (2001) 172404.

\bibitem{9} R. J. Radwanski and Z. Ropka, cond-mat/0211595.

\bibitem{10} J. H. Van Vleck, Rev. Mod. Phys. 50 (1978) 181.

\bibitem{11} R. J. Radwanski and Z. Ropka, \emph{Relativistic effects in the electronic structure for
the 3d paramagnetic ions}, cond-mat/9907140.

\bibitem{12} J. Spalek - the official rejection letter of 23.12. 2002 about our paper OO015PO. The Rector of the
Jagiellonian University in Krakow has been informed about this
scientific discrimination.

\bibitem{13} A. Abragam and B. Bleaney, \emph{Electron Paramagnetic
Resonance of Transition Ions} (Clarendon Press, Oxford) 1970 (pp
365-471).

\bibitem{14} C. J. Ballhausen, \emph{Ligand Field Theory} (McGraw-Hill
Comp.) 1962.

\bibitem{15} R. J. Radwanski and Z. Ropka, \emph{Influence of spin-orbit
interactions on the cubic crystal-field states of the d$^{4}$
system}, cond-mat/0201153.

\bibitem{16} S. Noguchi, S. Kawamata, K. Okuda, H. Nojiri, and M. Motokawa, Phys. Rev. B \textbf{66}
(2002) 094404.

\bibitem{17} Z. Ropka and R. J. Radwanski, Phys. Rev. B \textbf{67}
(2003) 172401.

\end{thebibliography}
\end{document}